\documentstyle[a4,11pt]{article}

\newcommand{\del}{\partial}
\newcommand{\Vka}{V_{\kappa}}
\newcommand{\Vla}{V_{\lambda}}
\newcommand{\Vmu}{V_{\mu}}

\newcommand{\Vkala}{V_{\kappa\lambda}}

\newcommand{\Vlaka}{V_{\lambda\kappa}}

\newcommand{\Vmuka}{V_{\mu\kappa}}

\newcommand{\Fkala}{F_{\kappa\lambda}}
\newcommand{\Fkanu}{F_{\kappa\nu}}
\newcommand{\Flaka}{F_{\lambda\kappa}}
\newcommand{\Flamu}{F_{\lambda\mu}}
\newcommand{\Fmunu}{F_{\mu\nu}}
\newcommand{\Fnumu}{F_{\nu\mu}}
\newcommand{\Fnuka}{F_{\nu\kappa}}
\newcommand{\Fmuka}{F_{\mu\kappa}}
\newcommand{\Fkalamu}{F_{\kappa\lambda\mu}}
\newcommand{\Flamunu}{F_{\lambda\mu\nu}}

\newcommand{\Fkamula}{F_{\kappa\mu\lambda}}
\newcommand{\Fkanumu}{F_{\kappa\nu\mu}}
\newcommand{\Fmulaka}{F_{\mu\lambda\kappa}}
\newcommand{\Fmulanu}{F_{\mu\lambda\nu}}
\newcommand{\Fmunuka}{F_{\mu\nu\kappa}}
\newcommand{\Fkalamunu}{F_{\kappa\lambda\mu\nu}}
\newcommand{\Flakanumu}{F_{\lambda\kappa\nu\mu}}

\begin{document}

\pagestyle{empty}

\renewcommand{\thefootnote}{\fnsymbol{footnote}}
\setlength{\parindent}{0pt}

\hskip 10.5cm {\sl HD-THEP-96/54}
\vskip.0pt
\hskip 10.5cm {\sl HUB-EP-96/57}
\vskip .7cm

\begin{center}

\vskip2cm
{\Large Application of the Worldline Path Integral Method\\}
{\Large to the Calculation of Inverse Mass Expansions
\footnote{ Talk given by D. Fliegner at AIHENP'96, Lausanne 
(Switzerland), Sept. 2-6 1996}

\vskip.7cm
{\large D. Fliegner$^{\rm \,\,a}$,
        P. Haberl$^{\rm \,\,b}$, 
        M. G. Schmidt$^{\rm \,\,a}$, 
        C. Schubert$^{\rm \,\,c}$ }
}

\vskip.7cm 

{\it $^{\rm a}$Institut f\"ur Theoretische Physik,
Universit\"at Heidelberg\\
Philosophenweg 16,
D-69120 Heidelberg, Germany}

\vskip.2cm

{\it $^{\rm b}$Institut f\"ur Theoretische Physik,
RWTH Aachen,\\
Sommerfeldstr. 26-28,
D-52056 Aachen, Germany}

\vskip.2cm

{\it $^{\rm c}$Institut f\"ur Physik,
Humboldt Universit\"at zu Berlin\\
Invalidenstr. 110, D-10115 Berlin, Germany}

\vskip1.4cm
 
{\large\bf Abstract}

\end{center}

\begin{quotation}
\noindent
Higher order coefficients of the inverse mass expansion of one--loop 
effective actions are obtained from a one--dimensional path integral
representation. For the case of a massive scalar loop in the background 
of both a scalar potential and a (non Abelian) gauge field explicit
results to ${\cal O}(T^5)$ in the proper time parameter are presented.
\end{quotation}

\vskip 1cm


\clearpage

\renewcommand{\thefootnote}{\protect\arabic{footnote}}

\pagestyle{plain}

\setcounter{page}{1}
\setcounter{footnote}{0}

\noindent
The inverse mass expansion of the one--loop effective action of a scalar 
particle with mass $m$ in the background of a (non Abelian) gauge field 
$A_\mu$ and a matrix valued scalar potential $V$ is usually obtained from 
the one--loop determinant of the corresponding fluctuation operator $M$,
which (in Euclidean space) reads

\begin{equation}
M = - D^2 + m^2 + V(x) \qquad{\rm with}\qquad
D_\mu=\del_\mu + i g  A_\mu (x)\,.
\end{equation}

\noindent
by using the Schwinger proper time representation,

\begin{equation}
\Gamma_{\rm eff}[A,V]=-{\rm ln}({\rm det M})=-{\rm Tr}({\rm ln} M)
=\int_0^\infty{{\rm d}T\over T}{\rm Tr}\;{\rm e}^{-T M}\;,
\end{equation}

\noindent
and expanding in powers of the proper time parameter $T$.  For the 
gauge case the coefficients of this expansion were calculated up 
to $O(T^5)$ by field theory methods in \cite{1}. The operator trace is 
nothing but the diagonal element of the heat kernel for the operator 
$M$, integrated over space-time. Using standard (recursive and 
non--recursive) heat kernel methods \cite{2} the coefficients have been 
computed up to $O(T^7)$ for the pure scalar case and $O(T^5)$ for the 
gauge case only recently \cite{3}. 

The worldline approach to the inverse mass expansion makes use of progress
in calculations of one--loop amplitudes by string inspired techniques 
\cite{4,5,6}.

The operator trace is written as a one-dimensional path integral over the 
space of closed loops in space-time with fixed proper time circumference 
$T$. The one--loop effective action takes the form

\begin{equation}
\Gamma[A,V]=
\int_0^\infty {{\rm d}T\over T} {\rm e}^{-m^2T} {\rm tr}
{\cal P} \int_{x(T)=x(0)} \hspace{-1cm} {\cal D}x
\exp \Bigl[ - \int_0^T {\rm d}\tau \bigl( {\dot{x}^2\over 4}
+ i g \, \dot{x}^\mu A_\mu  + V  \bigr)\Bigr]\,.
\end{equation}

\noindent
Here ${\cal P}$ denotes path ordering and the operator trace has reduced 
to an ordinary matrix trace. After expansion of the interaction exponential
the path integral can be evaluated by Wick contractions. The appropriate
Green function is the Green function of the Laplacian on the circle with
periodic boundary conditions \cite{5}. Since the defining equation of this 
Green function has no solution, one is forced to introduce a `background
charge' $\rho$ on the worldline. A very convenient choice is a uniformly 
distributed background charge $\rho=1/T$ (which distinguishes the present
formalism from former path integral approaches in this context \cite{7}), 
yielding the translationally invariant Green function

\begin{equation}
G_B(\tau_1,\tau_2) = |\tau_1-\tau_2|-{(\tau_1-\tau_2)^2\over T}
\;.
\end{equation}

\noindent
Due to the existence of the zero mode this Green function cannot be 
applied to the path integral as it stands. One has to introduce a loop
center of mass $x_0$ and a relative coordinate $y$,

\begin{equation}
x^{\mu}(\tau) = x^\mu_0 + y^\mu(\tau)
\qquad\mbox{with}\qquad
\int_0^T {\rm d}\tau \;y^{\mu}(\tau) = 0\,,
\end{equation}

\noindent
and separate the integration over the center of mass from the path integral,

\begin{equation}
\int {\cal D}x = \int {\rm d}^dx_0 \int {\cal D}y \;.
\end{equation}

\noindent
The resulting path integral is over the space of all loops with a common
center of mass $x_0$. Using $\dot{y}^\mu=\dot{x}^\mu$ this yields

\begin{equation}
\Gamma[A,V]=
\int_0^\infty {{\rm d}T\over T} {\rm e}^{-m^2T} {\rm tr} \;\;
{\cal P} \int \! \mbox{d}^d x_0 \int \! {\cal D}y \exp \Bigl[ - \int_0^T 
\!\!\!{\rm d}\tau \bigl( {\dot{y}^2\over 4} \!+\! i g \dot{y}^\mu A_\mu  
\!+\! V \bigr)\Bigr]\,. \,\,
\end{equation}

\noindent
If one chooses Fock-Schwinger gauge for the background gauge field with
reference to the center of mass $x_0$ \cite{6}, it can be written as

\begin{equation}
A_\mu(x_0+y) = y^\rho \int_0^1 {\rm d}\eta \;\eta \;
F_{\rho\mu}(x_0+\eta y)\;
\label{Acov}
\end{equation}

\noindent
and the background fields $F$ and $V$ can be expanded covariantly:

\begin{equation}
F_{\rho\mu}(x_0+\eta y) = {\rm e}^{\eta y D} F_{\rho\mu}(x_0) \,,\,\,\,\,
V(x_0+ y) = {\rm e}^{ y D} V(x_0) \;.
\label{Fcov}
\end{equation}

\indent
Using these formulas the expansion of the interaction exponentials in 
Eq.~(3) in the proper time parameter $T$ yields the following manifestly 
covariant form of the action:

\begin{eqnarray}
\Gamma[F,V] \!\!\!\!\! &=& \!\!\!\!\!\!\int_0^\infty \!{{\rm d}T\over T}
{\rm e}^{-m^2 T} {\rm tr}\int \! {\rm d}^d x_0 \sum_{n=0}^\infty
\frac{(-1)^n}{n} \int \! {\cal D}y \;{\rm exp} \Bigl[ -
\int_0^T \! {\rm d}\tau {\dot{y}^2\over 4} \Bigr]
\int_0^{\tau_1 = T} \hspace{-0.7cm}{\rm d}\tau_2 \ldots 
\!\int_0^{\tau_{n-1}} \hspace{-0.7cm}{\rm d}\tau_n
\nonumber\\ \!\!\!\!\!
&& \!\!\!\!\!\!\!\!\!\prod_{j=1}^n \!\Bigl[\;{\rm e}^{ y(\tau_j)D_{(j)}} 
V^{(j)}(x_0)\!+\! i g \dot{y}^{\mu_j}(\tau_j) y^{\rho_j}(\tau_j)\int_0^1 
\!\!\!\! {\rm d}\eta_j \eta_j {\rm e}^{\eta_j y(\tau_j) D_{(j)}}
F_{\rho_j\mu_j}^{(j)}(x_0)  \Bigr].
\label{master}
\end{eqnarray}

\noindent
The factor $\frac{1}{n}$ arises upon eliminating of the first
$\tau$-integration, which is possible due to the freedom of choosing the
origin somewhere on the loop.

\noindent
For the calculation of the heat kernel coefficients to a given order in $T$
one has to perform the following steps:

\begin{itemize}

\item[$\bullet$]{\em Wick contractions}

\noindent
The sum in Eq.~(10) is truncated and the exponentials are expanded. Then
all possible Wick contractions have to be evaluated using the contraction
rules

\begin{eqnarray}
\langle y^{\mu}(\tau_1) y^{\nu}(\tau_2) \rangle &=&
-g^{\mu\nu} G_B(\tau_1,\tau_2) =
-g^{\mu\nu} \Bigl[|\tau_1-\tau_2|-{(\tau_1-\tau_2)^2\over T}
\Bigr] \;,\nonumber\\
\langle \dot{y}^{\mu}(\tau_1) y^{\nu}(\tau_2)
\rangle &=& -g^{\mu\nu} \dot{G}_B(\tau_1,\tau_2) = -g^{\mu\nu}
\Bigl[ {\rm sign}(\tau_1-\tau_2)-{2(\tau_1-\tau_2)\over T}
\Bigr] \;,\nonumber\\
\langle \dot{y}^{\mu}(\tau_1) \dot{y}^{\nu}(\tau_2) \rangle &=&
+g^{\mu\nu} \ddot{G}_B(\tau_1,\tau_2) =
+g^{\mu\nu} \Bigl[2\delta(\tau_1-\tau_2)-{2\over T}\Bigr] \;,
\end{eqnarray}

\noindent
where the dot denotes differentiation with respect to the first 
variable. From these rules one may alternatively derive contraction 
rules for exponentials, which can be used equivalently. 

\item[$\bullet$]{\em Integrations}

\noindent
The polynomial $\tau$- and $\eta$-integrations are performed. After the
expansion in the first step the $\eta$-integrations are trivial.
The integrands of the $\tau$-integration involve the worldline Green 
function $G_B$ and its derivatives. They are polynomial in the 
variables $\tau_j$. 

\item[$\bullet$]{\em Basis reduction}

\noindent
The number of terms will be drastically reduced by an appropriate basis 
reduction, which is essential for the practical use of the coefficients
in numerical calculations.  In the pure scalar case there is a unique 
basis with the property that no box operators $\Box=\del^2$ occur. 
With the choice of the Green function Eq.~(4) no self contractions exist, 
i.e. box operators do never occur and no partial integrations have to be 
performed to reduce the result.  The only operation left is the 
identification of cyclic equivalent terms,  which is trivial. The gauge 
case is much more complicated and requires a reduction algorithm, which 
is described below. 

\end{itemize}

\noindent
Finally one obtains the result in the form

\begin{equation}
\Gamma[F,V] =
\int_0^\infty {{\rm d}T\over T} [4\pi T]^{-d/2} {\rm e}^{-m^2T}
{\sum_{n=1}^\infty} {(-T)^n\over n!} \!\!\int {\rm d}x_0
\;{\rm tr}\; O_n\;.
\end{equation}

\noindent
The factor $[4\pi T]^{-d/2}$ arises from the normalization of the free
path integral. For the scalar case the calculation of the coefficients 
was done up to $O_{11}$ \cite{8,9}.

In the following we describe the basis reduction algorithm for the gauge
case as proposed by M\"uller \cite{10}. It turns out that there is a minimal
basis of invariants without box operators and partial integrations are still 
not necessary to reduce our results into this basis. However, besides the 
cyclic permutations one has to use the antisymmetry of the field strength 
tensor, the Bianchi identities and the exchange of covariant derivatives, 
\begin{equation}
D_\mu D_\nu X = D_\nu D_\mu X + i g \, [F_{\mu\nu}, X ] \,, \mbox{etc.}
\end{equation}

for a further reduction.

\noindent
This is done by the following steps (for a more detailed discussion of
the algorithm and a proof of minimality see \cite{10}):

\noindent
\begin{itemize}

\item[$\bullet$]{\em Elimination of `middle' derivatives}

\noindent
The contractions of derivatives of a field strength tensor belong to 
different classes with respect to the contractions of the field strength
tensor itself. This can be seen very easily in a diagrammatic picture, 
where the (cyclic) function $tr$ is represented by a circle, the
fields by points on the circle and the contractions between the fields
by lines crossing the circle. In this picture the indices of a field 
strength tensor divide the circle into three sectors (left,middle,right). 
Consequently the derivatives fall into different classes, according to 
the sector, they are contracted with. The Bianchi identity involves 
all classes of derivatives and can be used to eliminate one of them. 
It is useful to take the symmetric choice and eliminate the middle
derivatives. This has to be done with all derivatives of a field strength
tensor, which possibly requires exchange of derivatives. 

\item[$\bullet$]{\em Reduction of multiple contractions between factors}

\noindent
Terms with multiple contractions are reduced by the following rules

\begin{eqnarray}
\mbox{tr} ( \ldots F_{\mu\nu} \ldots D_\mu F_{\nu\kappa} \ldots) & = &
\, \frac{1}{2} \mbox{tr} ( \ldots F_{\mu\nu} \ldots 
                               D_\kappa F_{\nu\mu} \ldots )\,,\\ 
\mbox{tr} ( \ldots D_\mu F_{\nu\kappa} \ldots 
                   D_{\nu} F_{\mu\lambda} \ldots ) & = &
\, \frac{1}{2} \mbox{tr}( \ldots D_\kappa F_{\nu\mu} \ldots 
                              D_{\lambda} F_{\mu\nu} \ldots ) \nonumber \\ 
          && + \, \mbox{tr} ( \ldots D_\mu F_{\nu\kappa} \ldots 
                         D_{\mu} F_{\nu\lambda} \ldots )\,,\\ 
\mbox{tr} ( \ldots F_{\mu\nu} \ldots D_\mu D_\nu X \ldots ) & = &
\frac{1}{2} \,i g\, \mbox{tr} ( \ldots F_{\mu\nu} \ldots [F_{\mu\nu}, X] ) \,,
\end{eqnarray} 

\noindent
which involve the Bianchi identity, the antisymmetry of the field strength
tensor and the exchange of derivatives. The aim of these rules is to produce 
contractions of the field strength tensors among themselves whenever 
possible. 

\item[$\bullet$]{\em Final arrangement of indices and factors}

\noindent
The indices still have to be ordered using the antisymmetry of the field 
strength tensor and the exchange of derivatives using e.g.~a minimal 
contraction distance rule. After this the cyclic equivalent invariants are
reduced as in the pure scalar case. 

\end{itemize}

\noindent
The reduction algorithm produces invariants with smaller number of
derivatives and higher number of field strength tensors by exchange 
of derivatives. Therefore one has to start the reduction with the terms
containing a maximum number of derivatives. 

For a more specialized situation one can use existing additional symmetries,
like the mirror symmetry \cite{10}, for a further reduction. For the general
case Table \ref{table} gives an overview of the number of invariants in the 
minimal basis to given order and number $v$ of occurring scalar background 
fields.

\begin{table}[h]
\begin{center}
\begin{tabular}{|c|r|r|r|r|r|r|r|r|} 
\hline
order & total & $v=0$ & $v=1$ & $v=2$ & $v=3$ & $v=4$ & $v=5$ & $v=6$\\
\hline
1 & 1	& 0	& 1	&  	&	&	&	&\\
2 & 2	& 1	& 0	& 1	&	&	&	&\\	
3 & 5	& 2	& 1	& 1	& 1	&	&	&\\
4 & 18	& 7	& 5	& 4	& 1	& 1	&	&\\
5 & 105	& 36	& 36	& 23	& 7	& 2	& 1	&\\ 
6 & 902	& 300	& 329	& 191	& 63	& 16	& 2	&1\\
\hline
\end{tabular}
\end{center}
\caption{\label{table} Number of basis invariants in different orders of 
the expansion.}
\end{table}

\noindent
The computerization of the calculation splits into two parts. The first
consists of the expansion of the interaction exponentials and the Wick 
contractions. As in the scalar case this is conveniently done with FORM 
\cite{11} up to ${\cal O}(T^6)$. The second part of the calculation consists 
of performing the $\tau$-integrals and the application of the basis algorithm 
to the raw coefficients. The coding of the basis algorithm requires a rule 
based system for symbolic manipulation with fast and flexible pattern 
matching.
For this purpose we chose the new language `M' developed by P. Overmann 
\cite{12}. The reduction has been completed for the coefficients up to 
${\cal O}(T^5)$, the reduction of $O_6$ is currently done. 
Up to ${\cal O}(T^5)$ the results have been checked to be equivalent with 
the result obtained from a modified non--recursive heat kernel method 
\cite{13}. Additionally, a check with the results of \cite{1} has been 
done up to ${\cal O}(T^4)$.

\noindent
The expression for $O_5$ is too large to be presented here and can be found
in \cite{9}. The results to ${\cal O}(T^4)$ read (absorbing the coupling 
constant $g$ into the fields, $\Fkalamunu\equiv D_\kappa D_\lambda 
F_{\mu\nu}$ etc.):

\begin{eqnarray*}
O_1 &=& V \,,\\
O_2 &=& V^2 + {1\over 6} \Fkala \Flaka \,,\\
O_3 &=& V^3 + {1\over 2} \Vka \Vka 
          + {1\over 2} V \Fkala \Flaka 
          - {2\over 15} \,i\,\Fkala\Flamu\Fmuka
          + {1\over 20} \Fkalamu\Fkamula \,,\\
O_4 &=& V^4 + 2 V\Vka\Vka + {1\over 5} \Vkala\Vlaka
          + {3\over 5} V^2 \Fkala\Flaka 
          + {2\over 5} V\Fkala V\Flaka \\
    &-& {4\over 5} \,i\, \Fkala\Vla\Vka 
          - {8\over 15} \,i\, V\Fkala\Flamu\Fmuka 
          + {1\over 5} V\Fkalamu\Fkamula 
          - {2\over 15} \Fkala\Flamu\Vmuka \\ 
    &+& {1\over 3}\Fkala\Fmulaka\Vmu 
          + {1\over 3} \Fkala\Vmu\Fmulaka 
          + {2\over 35} \Fkala\Flaka\Fmunu\Fnumu 
          + {4 \over 35} \Fkala\Flamu\Fkanu\Fnumu \\
    &-& {1\over 21} \Fkala\Flamu\Fmunu\Fnuka 
          - {8\over 105} \,i\,\Fkala\Flamunu\Fkanumu 
          - {6\over 35} \,i\, \Fkala\Fmulanu\Fmunuka \\
    &+& {11\over 420} \Fkala\Fmunu\Flaka\Fnumu  
          + {1\over 70} \Fkalamunu\Flakanumu \,.
\end{eqnarray*}

\noindent
In conclusion, we have obtained the inverse mass expansion of the one--loop
effective action for a massive scalar particle in the background of a non
Abelian gauge field and a matrix valued scalar potential up to 
${\cal O}(T^5)$, reduced to a minimal set of unique basis invariants. 
Within the framework of the worldline path integral method it is currently
feasible to compute the coefficient $O_6$. Since the basis reduction of this 
coefficient has not been completed yet, the results will be presented in a 
forthcoming publication \cite{13}.

\newpage


\newpage

\begin{thebibliography}{9}
\bibitem{1}
A.\ van de Ven, {\em Nucl.\ Phys.} {\bf B250} (1985) 593.
\bibitem{2}
R.\ D.\  Ball, {\em Phys.\ Rep.} {\bf 182} (1989) 1.
\bibitem{3}
A.\ A.\ Bel'kov, A.\ V.\ Lanyov, A.\ Schaale, {\em Comp.\ Phys.\
Commun.} {\bf 95} (1996) 123.
\bibitem{4}
Z.\ Bern, D.\ A.\ Kosower, {\em Phys.\ Rev.\ Lett.} {\bf B66} (1991) 1669;
{\em Nucl.\ Phys.} {\bf B379} (1992) 451;
Z.\ Bern, D.\ C.\ Dunbar, {\em Nucl.\ Phys.} {\bf B379} (1992) 562.
\bibitem{5}
M.\ J.\ Strassler, {\em Nucl.\ Phys.} {\bf B385} (1992) 145.
\bibitem{6}
M.\ G.\ Schmidt, C.\ Schubert, {\em Phys.\ Lett.} {\bf B318} (1993) 438.
\bibitem{7} E.\ Onofri, {\em Am. J. Phys.} {\bf 46} (1978) 444; Y.\ Fujiwara,
T.\ A.\ Osborn, S.\ F.\ Wilk, {\em Phys.Rev.} {\bf A 25} (1982) 14;
J.\ A.\ Zuk, {\em Nucl. Phys.} {\bf 280} (1987) 125.
\bibitem{8}
D.\ Fliegner, M.\ G.\ Schmidt, C.\ Schubert, {\em Z.\ Phys.}
    {\bf C64} (1994) 111;\\
D.\ Fliegner, P.\ Haberl, M.\ G.\ Schmidt, C.\ Schubert,
{\em Discourses in Mathematics and its Applications}, No. 4, 
(Texas A\&M,1995) p.87; 
{\em New Computing Techniques in Physics Research IV}, 
(World Scientific,Singapore,1995) p.199. 
\bibitem{9} available at
{\tt http://www.thphys.uni-heidelberg.de/$\tilde{\phantom{x}}$fliegner}
\bibitem{10}
U.\ M\"uller, {\em New Computing Techniques in Physics Research IV}
(World Scientific,Singapore,1995) p.193; DESY--96--154.
\bibitem{11} J.\ A.\ M.\ Vermaseren, {\em Symbolic Manipulation with
FORM} (CAN,1991).
\bibitem{12} P.\ Overmann, {\em The M Programming Language} (1996).
\bibitem{13} D.\ Fliegner, P.\ Haberl, M.\ G.\ Schmidt, C.\ Schubert, 
in preparation.
\end{thebibliography}
\end{document}